\documentclass[pre,twocolumn,showpacs,aps]{revtex4}
\usepackage{epsfig}
\newcommand{\av}[1]{\left<#1\right>}
\newcommand{\degrees}{$^{\circ}$}

\begin{document}
\title{Statistics of Velocity Gradients in\\
Two-Dimensional Navier-Stokes and Ocean Turbulence}
\author{
Norbert Schorghofer\thanks{E-mail: norbert@segovia.mit.edu}}
\affiliation{
Department of Earth, Atmospheric, and Planetary Sciences,
Massachusetts Institute of Technology, Cambridge, MA 02139}
\author{Sarah T. Gille}
\affiliation{Scripps Institution of Oceanography and
Department of Mechanical and Aerospace Engineering,
University of California San Diego, La Jolla, CA  92093-0230}
\date{\today}

\begin{abstract}
Probability density functions and conditional averages of velocity
gradients derived from upper ocean observations are compared with
results from forced simulations of the two-dimensional Navier-Stokes
equations.  Ocean data are derived from TOPEX satellite altimeter
measurements.  The simulations use rapid forcing on large scales,
characteristic of surface winds.  The probability distributions of
transverse velocity derivatives from the ocean observations agree with
the forced simulations, though they differ from unforced simulations
reported elsewhere.  The distribution and cross-correlation of
velocity derivatives provide clear evidence that large coherent eddies
play only a minor role in generating the observed statistics.
\end{abstract}

\pacs{47.27.Eq,92.10.Fj,92.10.Lq}
\maketitle

\section{Introduction}

\noindent
Statistical properties of turbulent flows, such as probability density
functions (pdfs), are important for characterizing turbulence.  For
instance, velocity gradients are directly related to velocity
correlations, relative dispersion, and energy dissipation in the fluid
\cite{monin}.  This study evaluates statistics of turbulence, 
as observed in recent satellite measurements
of the upper ocean.  Statistics of observed phenomena are compared
with corresponding statistics for the forced two-dimensional
Navier-Stokes equations.  Our results show that, in comparison with
unforced decaying turbulence, simple forced two-dimensional
Navier-Stokes equations provide better agreement with ocean
observations.

For this analysis, ocean velocities are derived from altimeter data
collected by the TOPEX/\linebreak[2]POSEIDON satellite, which performs
repeated measurements of the height $\eta$ of the ocean surface.  
We use only observations from the TOPEX altimeter,
which has lower noise levels than the POSEIDON instrument.  The
geostrophic relation, $v_x=(g/f)\partial \eta/\partial y$, yields the
velocity component perpendicular to the satellite ground track.
Surface geostrophic velocities are characteristic of sub-surface flow
in the ocean \cite{wun98}.  This geostrophic flow is typically
well-represented by two-dimensional shallow-water equations and
resembles two-dimensional turbulence \cite{rhi79,mcw84}.  The
derivative along the satellite track, $\partial_y v_x$, yields the
transverse velocity gradient.  We compute velocities
$v$ from consecutive high-frequency altimeter measurements
\cite{yale95,sstj96,lsg98,gls00}, and then determine velocity
gradients by computing along track differences over a distance of
12~km.  For comparison, the first baroclinic Rossby radius ranges between
10~km and 80~km between 60\degrees\ and 10\degrees\ latitude
\cite{houry87,stammer97}, so
transverse gradients over 12~km distance are expected to be
representative of mesoscale geostrophic motions.  The cross-track, or
longitudinal, derivative cannot be determined.  Higher order
derivatives are increasingly noisy.

Earlier results have shown that velocities typically have Gaussian
distributions within small regions of the ocean.  When satellite data
from the global ocean were combined, the resulting pdfs were
non-Gaussian, due to regional variations in velocity variance
\cite{lsg98,gls00}.  When velocities were normalized by their local
variances, the pdfs were Gaussian \cite{gls00}, at least for
well-sampled velocities within three standard deviations of the mean.
Similar results were obtained for subsurface floats deployed in the
North Atlantic Ocean, although analysis for velocities more than three
standard deviations from the mean indicated non-Gaussian tails
\cite{bra00a}.  The Lagrangian statistics of floats are however not
directly comparable to the results from the TOPEX altimeter, which
captures the Eulerian statistics.  In this study, we specifically
normalize velocities and velocity gradients by their local variances
before computing pdfs and other statistics.

We compare observed oceanic pdfs with simulations of two-dimensional
quasi-geostrophic flow.  The equations of motion are
\begin{equation}
\left( \frac{\partial}{\partial t} +
\frac{\partial \psi}{\partial y}\frac{\partial}{\partial x} -
\frac{\partial \psi}{\partial x}\frac{\partial}{\partial y}
\right) q = D \nabla^2 q + F,
\label{eq:eq}
\end{equation}
where the potential vorticity $q= -\nabla^2\psi + \psi/R^2$.  The 
second term in the potential vorticity is neglected both in the 
quasi-geostrophic limit of the shallow water equations and when
the Rossby radius $R$ is large in the homogeneous quasi-geostrophic
equations.  In either case, eq.~(\ref{eq:eq}) is equivalent to
two-dimensional Navier-Stokes flow.  In this study, we perform
simulations of the Navier-Stokes equations.  Rapidly varying
(white-in-time) forcing $F$ is applied on large scales, through random
stirring of the low vorticity modes.  This forcing resembles wind
forcing of the ocean, which varies rapidly in time but slowly in space
\cite{lar91,wik99}.  We consider an isotropic, homogeneous, and
statistically stationary state.  Simulations use a conventional
pseudo-spectral method and second-order dissipation.  Results were
obtained on a $1024\times1024$ grid with long time averaging.
Large-scale coherent vortices are clearly visible.  Further details
about the numerics and resulting velocity pdfs are described elsewhere
\cite{sch00a,sch00b}.

In many instances the velocity pdf is approximately Gaussian
\cite{frisch}, and this is also the case for the simulated turbulence
here \cite{sch00a}.  Far more conclusive than the velocity
distribution turns out to be the statistics of velocity derivatives.
A number of authors have investigated the velocity gradients of
three-dimensional Navier-Stokes turbulence
\cite{km89,kra90,vm91,yk91,fs91,bbpvv91,cdks93,bmtw97,sza97}, but here
we consider the far less studied two-dimensional case.  Measurements
of the transverse velocity derivatives are presented in
section~\ref{sec:globalpdf}.  Section~\ref{sec:pointvortex} discusses
evidence that large eddies alone provide only a minor contribution to
the observed statistics.  Section~\ref{sec:transandlong} briefly
discusses pertinent differences between forced and unforced
turbulence.  The last section contains conclusions.

\section{The probability distribution of velocity derivatives}
\label{sec:globalpdf}

Earlier work based on satellite altimeter data reported transverse
velocity gradient pdfs in small parts of the ocean \cite{lsg98}.
These results differed from gradient pdfs derived for decaying
two-dimensional turbulence, which showed an approximate Cauchy
distribution during the late stage of the evolution
\cite{jim96,mml96}.  The discrepancy is resolved by comparing to
simulations of stationary turbulence.

\begin{figure}
(a)\epsfig{file=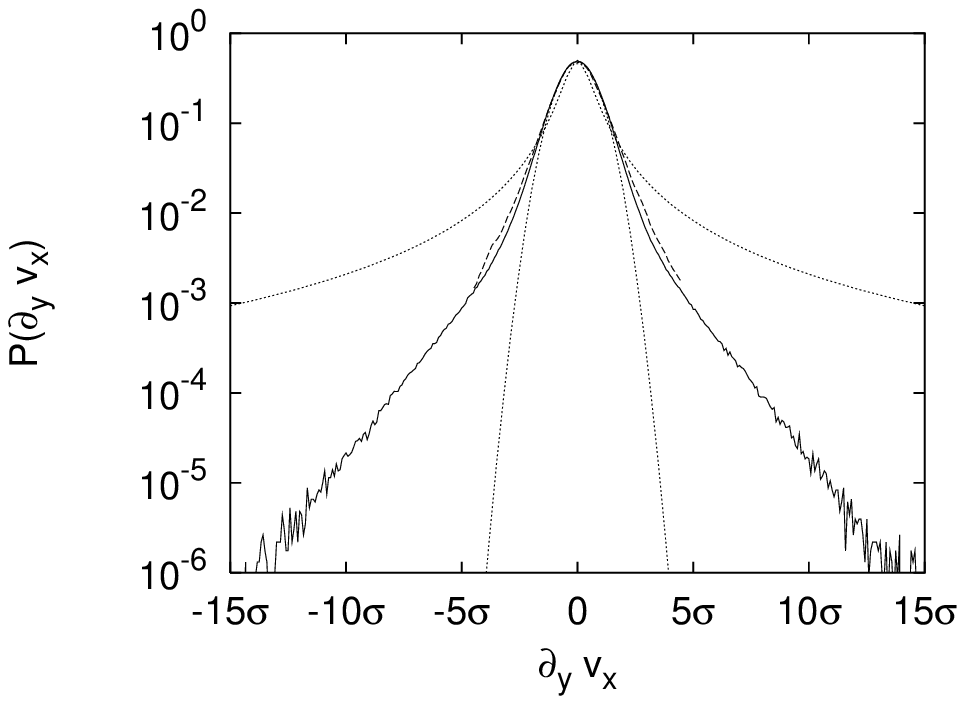,width=80mm}
(b)\epsfig{file=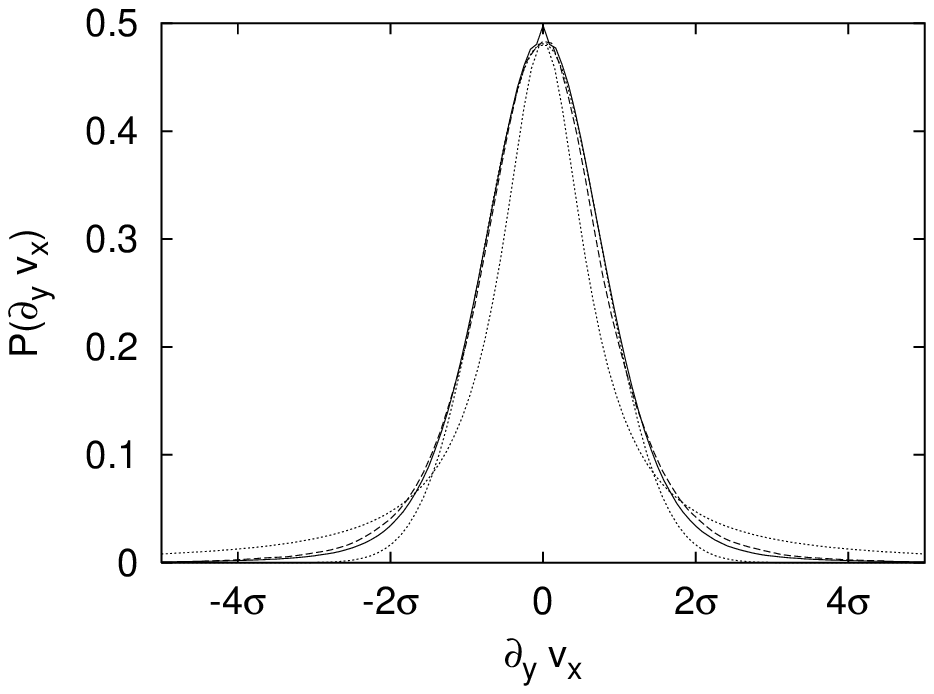,width=80mm}
\caption{
(a) Global variance-normalized pdf of the velocity gradient in the ocean 
(solid line) compared with simulations of two-dimensional Navier-Stokes 
turbulence (dashed line) on a semi-logarithmic scale.  
(b) Same quantities on a linear scale.  
In both cases, a Gaussian and a Cauchy distribution are shown for 
comparison (dotted lines).  
The ocean pdf is averaged over about 13 million data points.
Data are normalized by the standard deviation $\sigma$, as described 
in the text.
\label{fig:comparison}}
\end{figure}

Figure~\ref{fig:comparison} shows velocity gradient pdfs derived from 
ocean observations and simulations.  The solid line indicates the pdf 
of normalized velocity gradient data derived from global satellite 
altimetry.  To determine the oceanic pdf, velocity gradient data 
drawn from latitudes between 10\degrees\ and 60\degrees N and between 
10\degrees\ and 60\degrees S are sorted geographically into 
2.5\degrees\ by 2.5\degrees\ boxes.  Data near the equator are 
omitted because the geostrophic relationship is weak at low latitudes.  
The standard deviation of gradients in each latitude-longitude 
box varies from $1.4 \times 10^{-5}$~s$^{-1}$ near 60\degrees\ 
latitude up to $6.4 \times 10^{-5}$~s$^{-1}$ near 10\degrees\ latitude, 
with a median value of $1.9 \times 10^{-5}$~s$^{-1}$.  To compensate 
for this geographic variation, gradients are normalized to have 
unit standard deviation in each box, and then the pdf is computed 
from all of the normalized gradient data.  For comparison, we also 
normalized our pdfs using the mean absolute value of the velocity 
gradient; this did not diminish the strong tails of the gradient pdf.

The dashed line in Fig.~\ref{fig:comparison} represents the transverse
velocity gradient pdf from two-dimensional Navier-Stokes turbulence.
The dotted lines represent the narrow Gaussian distribution and the
broader Cauchy distribution, $P(x)=c/[\pi (c^2+x^2)]$, with long
tails.  The tails contribute noticeably to the standard deviation of
the pdf, and therefore the Gaussian is fitted to the data without
requiring unit area and unit standard deviation.  This is necessary to
make the Gaussian closely approximate the central part of the pdf.
Since the Cauchy distribution cannot be normalized by its standard
deviation, the constant $c$ is chosen such that $P(0)$ matches.  Both
the simulated and observed gradient pdfs appear Gaussian for small
velocity gradients, up to about one standard deviation.  For large
gradients they decay significantly more slowly than do Gaussian tails
but substantially faster than the Cauchy distribution found in
simulations of decaying turbulence \cite{jim96,mml96}.  There is good
agreement between observed and simulated pdfs up to even the largest
fluctuations measured in the simulation.

Error bars for the pdfs were estimated by grouping the data into $N$
groups and computing pdfs for each group.  Error of the mean pdf is
then taken to be the standard deviation of the pdf divided by
$\sqrt{N}$.  For this analysis, $N$ was the total number of
2.5\degrees\ boxes for the surveyed ocean or the number of velocity
snapshots for the simulation.  Since many ocean observations are
available, statistical errors are expected to be small compared to
systematic errors.  In fact, the statistical errors are frequently
narrower than the line width in Fig.~\ref{fig:comparison}.
Differences between the two distributions exceed the statistical
errors and are likely to be due to a number of factors, including
instrumental and atmospheric correction errors in the altimeter data,
which make the measurements noisy, as well as differences in the
physics of two-dimensional Navier--Stokes equations compared with the
ocean. 

The kurtosis (flatness), $\av{x^4}/\av{x^2}^2$, can serve as
quantitative comparison of the shape of the pdf.  In the simulation
results, the velocity gradient pdf has a kurtosis of $4.7$, indicating
clear deviation from Gaussian distribution.  If the observed ocean pdf
is terminated beyond the extent of the simulated one, at about four
and a half standard deviations, its kurtosis is also $4.7$.  This
quantitative comparison confirms that the oceanic pdf is substantially
better matched by the simulation than by either of the two ideal
distributions. 

Velocity gradient pdfs depend on the spatial separation between
velocity measurements.  The velocity correlations between two points
decrease with distance, and velocities at points very far apart can be
assumed to be statistically independent.  The distribution of velocity
differences across very large distances reduces to that of the
velocity (with twice the variance).  The 12 km separation of TOPEX
observations is small compared with the decorrelation length scales of
wind forcing, O(1000~km), and of mesoscale ocean features O(100~km),
so velocities at adjacent observation points are expected to be
strongly correlated.  Therefore, to obtain comparable results from the
numerical simulation, we have computed gradient pdfs from velocity
differences over asymptotically small separations.  For comparison, if
we compute gradient pdfs over progressively larger distances in the
simulation, then the distribution narrows from its original shape
(dashed line in Fig.~\ref{fig:comparison}) and becomes close to
Gaussian.

The basic simulations had a large-scale Reynolds number on the order
of $10^4$, while for ocean turbulence a Reynolds number of $10^7$
might be typical \cite{mel96}.  Pdfs were also determined for
simulations with lower and higher Reynolds numbers, using respectively
lower and higher resolutions, but shorter sampling time.  There is no
significant change in the shape of the pdfs \cite{tg96}, although
these data do not exclude a weak dependence on Reynolds number.  The
absence of any detectable Reynolds number dependence suggests that the
simulation data are close to what they look like at substantially
higher Reynolds number.  The difference in the length of the tails in
Fig.~\ref{fig:comparison} may be due to the vast difference in
Reynolds number, difference in sampling size, and errors from the
numerical differentiation of data.

The real ocean differs from the forced Navier--Stokes system because
of the addition of the $\beta$-effect, stratification, three-dimensional
motions, and buoyancy.  Hence it is surprising that there is such a
close agreement between measurement and simulation. In any case, the
agreement between observation and simulation suggests that the oceanic
velocity statistics may be understood within the framework of
two-dimensional turbulence.

\section{The role of coherent vortices}
\label{sec:pointvortex}

Idealized models of point vortices predict a Cauchy distribution for
the velocity gradients and a Gaussian distribution for the velocity
\cite{jim96,mml96,kus00,cha00a,chavanis01}.  This agrees with results
from decaying two-dimensional turbulence \cite{jim96,arr95}.  Hence,
in the late stages of decay, the statistics of velocity gradients have
been successfully understood to result from the far-field of
well-separated vortices \cite{jim96,mml96}.  In contrast, pdfs of
ocean surface velocity gradients are observed to have more rapidly
decaying tails than do Cauchy distributions.  Also as 
Fig.~\ref{fig:comparison}  makes evident, the simulations of
stationary two-dimensional turbulence show far less pronounced tails
than a Cauchy distribution.  

\begin{figure}
\epsfig{file=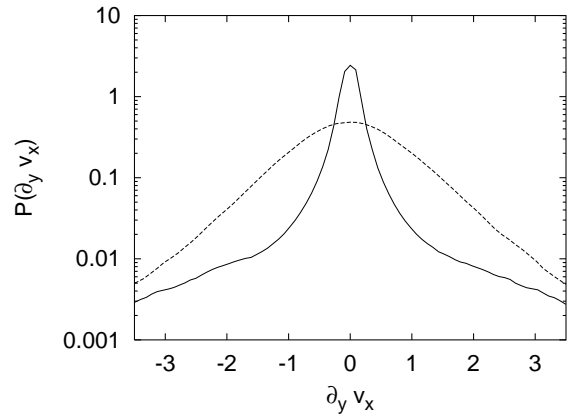,width=80mm}
\caption{Pdf of the velocity gradient in simulations of
two-dimensional Navier-Stokes turbulence.  The solid line is for the
velocity gradients produced by large coherent vortices.  The dashed
line corresponds to the complete flow field.  Both distributions are
normalized by the same standard deviation, hence preserving
differences in their width.
\label{fig:peak}}
\end{figure}

The discrepancy arises not only in the shape of the distribution but
also in its width.  A straight-forward way of illustrating this is to
calculate the velocity field produced by vortices with vorticities
that exceed twice the root-mean-square value.  Figure~\ref{fig:peak}
shows the transverse velocity gradients produced by these coherent
vortices (solid line).  For comparison, the actual distribution is
shown as a dashed line.  Clearly the large coherent vortices do not
generate enough intermediate gradients.  (Nor, for that matter, do
they account for most of the velocities.)  Consequently, the
distribution of gradients is poorly accounted for by large-scale
coherent vortices.

The contribution of the small-scale turbulence is also relevant.  This
agrees with the basic physical picture, according to which the late
stage of decaying turbulence consists of coherent vortices.  Its
statistics can therefore be understood in terms of them.  In the
stationary case, on the other hand, fluctuations over a wide spectrum
of spatial scales contribute to the gradient statistics.

Available statistical variables from the altimeter are the velocity
and the transverse velocity derivative.  Hence, one can study the
cross-correlation between these two quantities.  Here, we examine the
conditional average of the squared velocity gradient as a function of
velocity, $\av{(\partial_y v_x)^2 | \, |v_x|}$, which is the average
of $(\partial_y v_x)^2$ over all points with velocity component $\pm
v_x$.  The slope of $\av{(\partial_y v_x)^2 | \, |v_x|}$ is a measure
of the correlation between the velocity and the transverse velocity
derivative.  If there were no correlation between the velocity at a
point and the gradient at the same point, the conditional average
would be constant for all values of $v_x$ and would be exactly one if
velocity gradients were normalized by their standard deviation.  In
contrast, if gradients and velocities were strongly correlated, as 
would be expected around an isolated vortex, then the graph for the 
conditional average would have a pronounced slope.

\begin{figure}
\epsfig{file=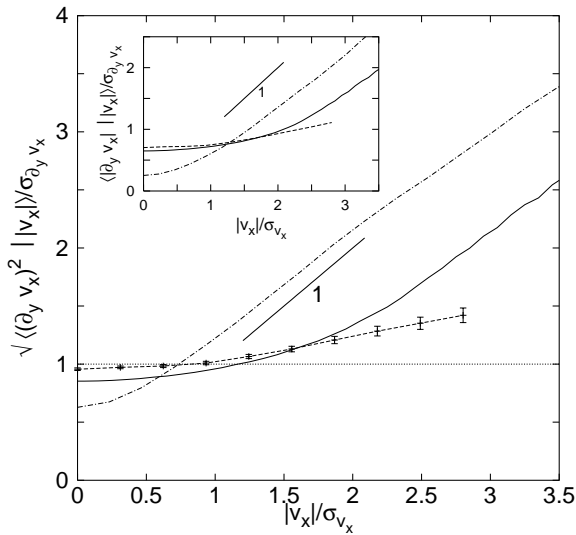,width=80mm}
\caption{Conditional average of transverse velocity gradient with
velocity, $\sqrt{\av{(\partial_y v_x)^2| \, |v_x|}}$, for
two-dimensional Navier-Stokes (dashed line with error bars) and ocean
turbulence (solid line).  The conditional average produced by the
large vortices in the simulation is also shown (dash-dotted line).  All
three graphs are normalized by their respective standard deviations.
The thick solid line indicates growth proportional to the velocity.
The error bars include only the statistical error expected from
averaging of the 32 snapshots, showing twice the standard error of the
mean.  The inset shows the conditional average of the absolute value $\av{|\partial_y v_x| \, | \, |v_x|}$.
\label{fig:cond_av}}
\end{figure}

Figure~\ref{fig:cond_av} shows the square-root of the measured
conditional average together with that for two-dimensional
Navier-Stokes turbulence.  For the ocean, we have normalized both
$v_x$ and $\partial_y v_x$ by their standard deviations in each
2.5\degrees\ by 2.5\degrees\ geographic box, because both quantities
vary spatially.  The axes are labeled in units of their respective
standard deviations, $\sqrt{\av{v_x^2}}$ and $\sqrt{\av{(\partial_y
v_x)^2}}$.  The correlation between velocity and its gradient is weak,
but the gradients tend to be higher when the velocity is large.  If
oceanic gradients beyond four standard deviations are excluded, which
is a fairer comparison with the simulation, the conditional average is
closer to one.  The longitudinal component of the conditional average
(not shown) exhibits behavior similar to the transverse component.
Also shown in Fig.~\ref{fig:cond_av} is the conditional average for
the velocity field of vortices larger than twice the root-mean-square
vorticity (dash-dotted line).  As expected there is a comparatively strong
correlation between velocities and velocity derivatives.  At large
velocities the slopes of the graphs for the coherent vortices and the
ocean are similar.  This may indicate influence by large eddies in
regions where the velocities are high, although no corresponding
evidence is found in the probability distribution of the gradients.
The situation at high velocities is therefore somewhat ambiguous.  For
small velocities, which cover most of the area, conditional averages
are near one for simulations (dashed line) and observations (solid
line), indicating that at low velocity, gradients are almost
uncorrelated with velocity.
The inset in Fig.~\ref{fig:cond_av} shows the conditional average
using $|\partial_y v_x|$ instead of $(\partial_y v_x)^2$, which is
less sensitive to outliers.  For small velocities, the agreement
between observation and simulation is closer and the discrepancy
between the large eddy field and the other two conditional averages
is stronger.  %For large velocities the gradients in the ocean
%still grow almost linear with velocity.

The deviation of the observed conditional averages from that for
coherent vortices strengthens the evidence that the gradient
statistics are unaccounted for by the velocity field created by
large-scale coherent eddies.  The role of coherent vortices in
generating the observed velocity statistics is minor.  This conclusion
cautions against attempts to model oceanic velocity fields by large
eddies.

\section{Forced versus unforced turbulence}
\label{sec:transandlong}

\begin{figure}
(a)\epsfig{file=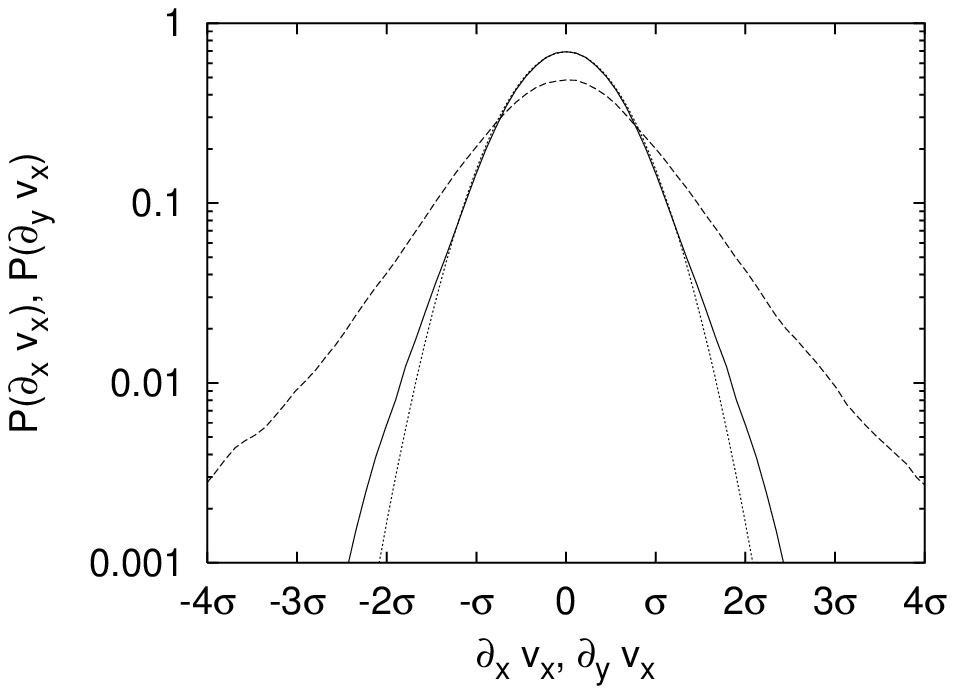,width=80mm}
(b)\epsfig{file=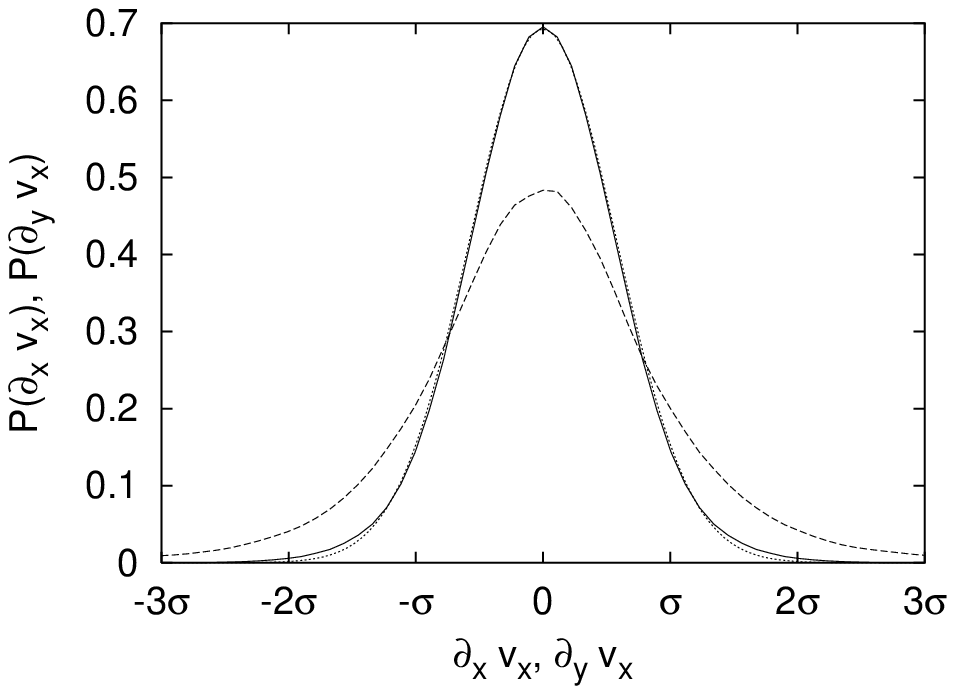,width=80mm}
\caption{
(a) Probability density functions of velocity derivatives for 
forced two-dimensional Navier-Stokes turbulence on a semi-logarithmic 
scale.  
(b) Same quantities on a linear scale.  
In both panels, the dashed line shows the transverse component, 
$\partial_y v_x$, and the solid line the longitudinal component, 
$\partial_x v_x$.  Both are normalized by the standard deviation of 
the transverse component $\sigma=\sqrt{\av{(\partial_y v_x)^2}}$.  
The dotted line is a Gaussian.  Transverse and longitudinal 
gradient pdf differ from each other in width and shape.
\label{fig:velgrad}}
\end{figure}

Although only the transverse velocity component can be determined from
altimeter data, simulations also permit us to examine the longitudinal
derivative, $\partial_x v_x$.  Figure~\ref{fig:velgrad} shows a clear
difference between the behavior of the longitudinal and transverse
components.  In our forced simulations, the standard deviation of
longitudinal fluctuations is about 60\% of that for transverse
fluctuations.  In isotropic and incompressible turbulence there is an
exact relation between the standard deviation of transverse and
longitudinal component \cite{batchelor}.  With a calculation analogous
to the well-known three-dimensional case \cite{batchelor}, we find in
the two-dimensional case $\av{(\partial_y v_x)^2}=3\av{(\partial_x
v_x)^2}$.  Hence, the standard deviation for the longitudinal
component is $1/\sqrt{3}\approx 58\%$ of that for the transverse
component.  This agrees with the measured value of 60\% in the
simulation.

Longitudinal and transverse pdfs differ not only by a factor of
$\sqrt{3}$ in their standard deviation, but also in their shape.
While the transverse gradients strongly deviate from a Gaussian
distribution, the longitudinal gradient pdf more closely approximates
a Gaussian.  The kurtosis of the longitudinal component is $3.5$,
substantially closer to the Gaussian value of $3$ than the transverse
component is, implying that large longitudinal gradients occur less
frequently than do large transverse gradients.  For simple
point-vortex models both, transverse and longitudinal components, are
distributed like Cauchy distributions (albeit with different standard
deviations) \cite{jim96}.  Also in the late stage of decaying
turbulence, the transverse component is distributed in the same way as
the longitudinal component \cite{jim96}.  This is yet another
difference between forced and unforced turbulence.

Overall, our study establishes a clear distinction between the
gradient statistics of unforced (freely decaying) and forced
(stationary) turbulence.  The presence of
forcing not only influences the properties of large-scale vortices but
also changes the distribution of eddies over different scales.
(Freely decaying turbulence has an inverse energy cascade while
two-dimensional turbulence forced at large scales is governed by a
direct enstrophy cascade.)  Further study is needed to determine 
how the statistics may depend on the temporal and spatial structure
of the forcing.

\section{Conclusions}

In conclusion, we find that transverse velocity derivative pdfs from
observed upper-ocean turbulence agree closely with forced
two-dimensional simulations but differ from previously reported
unforced turbulence.  The forcing diminishes the role of coherent
vortices in the pertinent statistics.  The distribution and
cross-correlation of velocity derivatives provide clear evidence that
large coherent eddies play only a minor role in generating the
observed statistics.  Further study of forced two-dimensional
turbulence appears likely to shed light on the character of meso-scale
turbulence in the ocean.

%\vspace{1em}\noindent
%{\bf Acknowledgments.}  
\begin{acknowledgments}
The work of N.S. was supported by a grant from the Research Grants 
Council of the Hong Kong Special Administrative Region, China 
(RGC Ref. No. CUHK4119/98P), by a postdoctoral fellowship from 
the Chinese University of Hong Kong, and by the Massachusetts 
Institute of Technology.
S.T.G. was supported by NASA through the Jason Altimeter Science 
Working Team (JPL contract 1204910).
\end{acknowledgments}

%\bibliographystyle{unsrt}
%\bibliographystyle{apsrev}
%\bibliography{/home/norbert/Papers/my} 

\end{document}